\documentclass[12pt]{article}

\usepackage{amsmath}
\usepackage{amsfonts}
\usepackage{amssymb}
\usepackage{graphicx}
\usepackage{psfrag}
\usepackage{graphicx}
\usepackage{psfrag}

\newcommand{\be}{\begin{equation}}
\newcommand{\ee}{\end{equation}}
\newcommand{\ba}{\begin{eqnarray}}
\newcommand{\ea}{\end{eqnarray}}

\begin{document}
\begin{center}
{\bf
GENERAL SOLUTION OF THE TWO-DIMENSIONAL INTERTWINING RELATIONS FOR SUPERCHARGES WITH HYPERBOLIC (LORENTZ) METRICS}\\
\vspace{1cm}
{\large \bf M. S. Bardavelidze$^{1,}$\footnote{E-mail: m.bardaveli@gmail.com},
M. V. Iof\/fe$^{2,}$\footnote{E-mail: m.ioffe@pobox.spbu.ru},
D. N. Nishnianidze}$^{1,2,}$\footnote{E-mail: cutaisi@yahoo.com}\\
\vspace{0.5cm}
$^1$ Akaki Tsereteli State University, 4600 Kutaisi, Georgia\\
$^2$ Saint-Petersburg State University,198504 Sankt-Petersburg, Russia
\end{center}
\vspace{0.5cm}
\hspace*{0.5in}
\vspace{1cm}
\hspace*{0.5in}
\begin{minipage}{5.0in}
{\small
The supersymmetrical intertwining relations are the most productive part of the supersymmetrical method in two-dimensional Quantum
Mechanics. Most interesting are relations with hyperbolic form of derivatives in supercharges. So far, several explicit solutions were found,
and they provide nontrivial two-dimensional potentials which were further studied by means of supersymmetrical methods. In the present paper
the general solution of intertwining relations with hyperbolic structure of supercharges is obtained. The corresponding potentials
are built explicitly, and it is evident that some of them were not known before now.}
\end{minipage}


\section*{\normalsize\bf 1. \quad Introduction.}

The method of Supersymmetrical Quantum Mechanics (SUSY QM) is known as effective method of studying one-dimensional \cite{review}, \cite{ai}
and two-dimensional quantum systems \cite{Ioffe}, \cite{Ioffe-2}, \cite{ai} in modern Quantum Mechanics. From practical point of view,
the most important relation of SUSY algebra is the vanishing commutator of superHamiltonian with supercharge operators. In components,
it corresponds to the so called supersymmetrical intertwining relations between two partner Hamiltonians of Schr\"odinger form. Many
generalizations of the standard SUSY algebra hold the form of intertwining relations but using different expressions for supercharges.

In particular, in the framework of two-dimensional SUSY QM \cite{TMF} the second order intertwining relations are:
\be
H_1Q^+=Q^+H_2, \label{int}
\ee
where the partner Hamiltonians
\be
H_i=-(\partial_1^2+\partial_2^2)+V_i(\vec{x}),\quad i=1,2, \label{ham}
\ee
and supercharges $Q^{\pm}$ can be taken with second order terms of hyperbolic (Lorentz) kind:
\be
Q^+=\partial_1^2-\partial_2^2+C_k(\vec{x})\partial_k+B(\vec{x}).\label{Q}
\ee
The coefficient functions $C_k(\vec x),\, B(\vec x)$ and potentials $V_i(\vec x)$ have to be found from (\ref{int}).
It was shown in \cite{TMF}, that these functions can be expressed in terms of functions $F_1(2x_1),\, F_2(2x_2),\, C_{\pm}(x_{\pm})$
as follows ($x_{\pm}\equiv x_1\pm x_2$):
\ba
&&2C_1(\vec{x})=C_+(x_+)+C_-(x_-),\quad 2C_2(\vec{x})=C_+(x_+)-C_-(x_-),\nonumber\\&&
V_{1,2}(\vec{x})=\pm\frac{1}{2}(C'_++C'_-)+\frac{1}{8}(C_+^2+C_-^2)+\frac{1}{4}\biggl(F_2(2x_2)-F_1(2x_1)\biggr),\label{V}\\&&
B(\vec{x})=\frac{1}{4}\biggl(C_+C_-+F_1(2x_1)+F_2(2x_2)\biggr).\label{B}
\ea

The intertwining relations (\ref{int}) are equivalent to a system of six nonlinear differential equations for $C_{\pm}, F_{1,2}, V_{1,2}.$
It was shown that this system can be written compactly as:
 \ba
 &&\partial_-(C_-F)=-\partial_+(C_+F), \quad F\equiv F_1(2x_1)+F_2(2x_2),\label{1}\\
 &&\partial_+^2F=\partial_-^2F.\label{2}
 \ea
The system (\ref{1}), (\ref{2}) is a system of functional-differential equations since each of functions $F_{1,2}, C_{\pm}$ depends
on its own parameter.

In the present paper, the intertwining relations for supercharges with hyperbolic metrics in the form of system (\ref{1}), (\ref{2})
will be analyzed. The general solution will be derived in Section 2, where seven options for the coefficient functions will be found.
It will be shown that some of them are already known. In Section 3, just the previously unknown partner potentials will be written out.

\section*{\normalsize\bf 2. \quad Solutions of intertwining relations.}

For the cases $C_-=0$ or $C_+=0$ the general solutions of (\ref{1}), (\ref{2}) were given in \cite{TMF}. Below we will investigate the case
of $C_+C_-\neq 0.$ Then, the general solution of (\ref{1}) can be written as:
\ba
 F=L(\int\frac{dx_+}{C_+}-\int\frac{dx_-}{C_-})/C_-C_+,\label{3}
 \ea
 where the function $L$ has to be found from (\ref{2}). It is convenient to define new unknown functions
 $A_{\pm}'\equiv 1/C_{\pm}(x_{\pm}),$ so that substitution of (\ref{3}) in (\ref{2}) gives new functional-differential
 equation for $L$ and $A_{\pm}:$
 \ba
 (\frac{{A_+}'''}{{A_+}'}-\frac{{A_-}'''}{{A_-}'})L(A_+-A_-)+3({A_+}''+
 {A_-}'')L'(A_+-A_-)+({A_+}'^2-{A_-}'^2)L''(A_+-A_-)=0.\label{4}
 \ea
It is possible to represent arbitrary $A_{\pm}$ formally as solutions of equations:
 \ba
 {A_+}'^2(x_+)=M_+(A_+),\quad {A_-}'^2(x_-)=M_-(A_-) \label{5}
 \ea
 with some functions $M_{\pm}.$ Then, (\ref{4}) takes the form:
 \ba
 &L(A_+-A_-)[{M_-}''(A_-)-{M_+}''(A_+)]+2L''(A_+-A_-)[M_-(A_-)-M_+(A_+)]\nonumber\\
 &-3L'(A_+-A_-)[{M_-}'(A_-)+{M_+}'(A_+)]=0.\label{6}
 \ea
This equation, which will play an important role below, coincides exactly, up to substitutions
$$x_1\to A_-; \quad x_2\to A_+; \quad V_1\to M_-; \quad V_2\to M_+; \quad V_{12}\to L(A_--A_+),$$
with the functional-differential equation, which was investigated
in \cite{inozem}, where its general solution was obtained (see Appendix). The representation (\ref{5})
 was necessary just to provide this interrelation.

Thus, the general solution of equation (\ref{6}) provides solution for function $L(A_--A_+)$
and also the ordinary differential equations for the corresponding functions $A_{\pm}.$
This result can be separated into seven different opportunities (variants):
 \ba
 I.\,\, L(A_+-A_-)=const,\quad A'^2_{\pm}(x_{\pm})=aA^2_{\pm}+bA_{\pm}+c;\label{7}
 \ea
 \ba
 II.&& A'^2_-(x_-)=a_2\exp(2\lambda A_-)+b_2\exp(\lambda
 A_-)+c,
 \nonumber\\&&A'^2_+(x_+)=a_1\exp(-2\lambda A_+)+b_1\exp(-\lambda
 A_+)+c,\, L(A_+-A_-)=a\exp(\lambda(A_+-A_-));\label{8}
 \ea
 \ba
 III.&&\quad {A_-'}^2(x_-)=a_1\delta^2\exp(-4\lambda A_-)-b_1\sigma^2\exp(4\lambda A_-)
 +a_2\delta\exp(-2\lambda A_-)-b_2\sigma\exp(2\lambda
 A_-)+k,\nonumber\\&&
 {A_+'}^2(x_+)=a_1\sigma^2\exp(-4\lambda A_+)-b_1\delta^2\exp(4\lambda A_+)
 +a_2\sigma\exp(-2\lambda A_+)-b_2\delta\exp(2\lambda
 A_+)+k,\nonumber\\&&
 L(A_--A_+)=\frac{a}{(\sigma\exp(\lambda (A_--A_+))-
 \delta\exp(-\lambda (A_--A_+)))^2};\label{9}
 \ea
 \ba
 IV.\quad {A_{\pm}'}^2(x_{\pm})=\sum_{k=0}^4a_kA_{\pm}^k,\quad
 L(A_--A_+)=\frac{n}{(A_--A_+)^2};\label{10}
 \ea
 \ba
 V.&&\quad {A_-}'^2(x_-)=a_1\sigma\exp(-2\lambda A_-)-b_1\delta\exp(2\lambda
 A_-)+c_1,\nonumber\\
 &&{A_+}'^2(x_+)=a_1\delta\exp(-2\lambda A_+)-b_1\sigma\exp(2\lambda
 A_+)+c_1,\nonumber\\&&
 L(A_--A_+)=\frac{\delta\exp(\lambda(A_--A_+))+\sigma\exp(-\lambda(A_--A_+))
 +\gamma}{(\delta\exp(\lambda(A_--A_+))-\sigma\exp(-\lambda(A_--A_+)))^2};
 \label{11}
 \ea
 \ba
 VI.\quad {A_-}'^2(x_-)=aA_-,\quad {A_+'}^2(x_+)=-aA_+,\quad
 L(A_--A_+)=b(A_--A_+)+c;\label{12}
 \ea
 \ba
 VII.\quad A_{\pm}'^2(x_{\pm})=aA_{\pm}^2+bA_{\pm}+c,\quad
 L(A_--A_+)=n+\frac{m}{(A_--A_+)^2}.\label{13}
 \ea
All letters above (besides $L$ and $A_{\pm}$) are arbitrary constants.

Some of variants I -- VII were already studied. In particular, if $F(\vec{x})$ is factorizable
$F(\vec{x})=F_+(x_+)F_-(x_-),$ the Eqs.(\ref{1}), (\ref{2}) can be directly easily solved.
These solutions were given in \cite{TMF}, \cite{Classical}, \cite{Ioffe}.
The function $F(\vec{x})$ is factorizable simultaneously with the function $L,$ as follows from (\ref{3}).
In terms of seven variants above, it means that solutions of variants I and II, and also of variant III with $\delta\sigma =0$
are already known.
Also, one can check that variants VI and VII with different values of constants were studied already in papers \cite{Classical}, \cite{Ioffe}.

In the variant III with $\delta\sigma\neq 0$, the functions $A_{\pm}$
are defined up to additional arbitrary constant. Therefore, one can make
$\delta =\sigma =1$ by means of suitable translation. After that, in terms of new functions
$U_{\pm}\equiv\exp(-2\lambda A_{\pm}),$ one has:
 \ba
 C_{\pm}=m\frac{U_{\pm}}{U_{\pm}'},\quad
 F=\frac{U_-'U_+'}{(U_+-U_-)^2},\quad {U_{\pm}'}^2=\sum_{n=0}^4\tilde{d}_n
 U_{\pm}^n\equiv P_4(U_{\pm}).\label{14}
 \ea
In turn, the variant IV
  \ba
  C_{\pm}=\frac{1}{A_{\pm}'},\quad F(\vec x)=\frac{A_-'A_+'}{(A_--A_+)^2},\quad
  {A_{\pm}'}^2=\sum_{k=0}^4a_kA_{\pm}^k, \label{15}
  \ea
can be reduced to a particular case of variant III. Indeed, replacing
$U_{\pm}\to n+mU_{\pm}$ in Eq.(\ref{14}) of variant III, one obtains:
 \ba
 C_{\pm}=\frac{n+mU_{\pm}}{U_{\pm}'},\quad
 F=\frac{U_-'U_+'}{(U_+-U_-)^2},\quad {U_{\pm}'}^2=\sum_{n=0}^4d_n
 U_{\pm}^n\equiv P_4(U_{\pm}).\label{16}
 \ea
which obviously includes variant IV (\ref{15}) (up to replacement $A_{\pm}\leftrightarrow U_{\pm}$).

In the variant V, two different cases i) $\delta\sigma\neq 0 $ and ii) $\delta\sigma=0$ must be considered separately. Again,
by suitable translations of functions $A_{\pm}$
one can restrict analysis of opportunity i) to the cases $\delta =\sigma =1$ or to $\delta=-\sigma=1,$ depending on the signs of initial
$\delta , \sigma .$ It is convenient to introduce new functions $U_{\pm}\equiv\exp(-2\lambda A_{\pm})$ for $\delta =\sigma=1,$ and
$U_{\pm}\equiv -\exp(-2\lambda A_{\pm})$ for $\delta =-\sigma =1$. Then, both choices of constants lead to the same expressions for $C_{\pm}$
and $F,$ and the same equations for functions $U_{\pm}$. Namely, the option i) reads:

i)\ba
  &&C_{\pm}=k\frac{U_{\pm}}{U'_{\pm}},\quad
  F=k_1\frac{U_+'U_-'}{(U_+-U_-)^2}+k_2\frac{U_+'U_-'(U_++U_-)}
  {\sqrt{U_+U_-}(U_+-U_-)^2},\nonumber\\&&
  U_{\pm}'^2=aU_{\pm}^3+bU_{\pm}^2+cU_{\pm}\equiv P_3(U_{\pm}),\label{17}
  \ea
where $k, k_1, k_2, a, b, c$ - arbitrary constants.
It should be noted that expressions for $C_{\pm}$ and $F$ are not changed under $U_{\pm} \leftrightarrow 1/U_{\pm}.$ It is
equivalent to $a \leftrightarrow c$ in (\ref{17}), i.e. the problem is symmetrical under $a\leftrightarrow c.$

Situations with different values of $a,b,c$ were already considered. For example, the case of $abc\neq 0$ and the polynomial $P_3(U_{\pm})$
with degenerate roots were considered in paper \cite{Classical}.
The particular case $a=0, bc\neq 0$ is equivalent to the case $ab\neq 0, c=0,$ and they were considered in papers \cite{IV},\cite{INV},\cite{IKN}.
The case with $a=b=0$ is equivalent to one with $a=c=0,$ and they give the potentials which are amenable to conventional separation of variables.
The case with $b=0, ac\neq 0$ corresponds to the situation without degeneration of roots for polynomial $P_3(U_{\pm}).$ This problem was studied in
\cite{IV}, \cite{IGV}, where the reducible
second order supercharges with twist were studied. In particular, it was shown in \cite{IGV} that the problem is reduced to solution of
a pair of nonlinear ordinary differential equations for functions $C_{\pm}(x_{\pm}):$
\be
{C'_{\pm}}^2=\alpha C_{\pm}^4+\beta C_{\pm}^2+\gamma.\label{redius}
\ee
with arbitrary constants $\alpha, \beta, \gamma .$ By the direct substitution of $C_{\pm}$ from (\ref{17}) into (\ref{redius}) one can check that
the latter equation is satisfied, and constants $\alpha, \beta, \gamma$ can be expressed in terms of constants $a, b, c, k$ of
(\ref{17}), i.e. the case (\ref{17}) corresponds to reducible supercharge with twist.

For the option ii) of variant V, a more convenient definition is $U_{\pm}\equiv\exp(\mp2\lambda A_{\pm}).$ Then,
for $\delta=0,\sigma\neq 0$ (analogously for $\delta\neq 0,\sigma =0 $) one has:

ii)
  \ba
  &&C_{\pm}=\pm p\frac{U_{\pm}}{U'_{\pm}},\quad F=n_1\frac{U'_-U'_+}
  {\sqrt{U_-U_+}}+n_2U'_-U'_+,\nonumber\\
  &&{U_-}'^2=dU_-^2+a_-U_-,\quad {U_+}'^2=dU_+^2+a_+U_+.\label{18}
  \ea
This case was already considered in papers \cite{Classical}, \cite{Ioffe}.

\section*{\normalsize\bf 3. \quad New potentials.}

Thus, there is a chance to obtain some new potentials only from Eq.(\ref{16}). It is useful to break the problem into several parts.

\textbf{1) Polynomial $P_4(U_{\pm})$ has degenerate roots,} i. e.
$U'^2_{\pm}=(U_{\pm}-u_0)^2(d_4U_{\pm}^2+n_1U_{\pm}+u_2).$
After replacement $U_{\pm}\to u_0 +1/Z_{\pm},$ Eq.(\ref{16}) can be rewritten as:
 \ba
 C_{\pm}=-\frac{(n+mu_0)Z^2_{\pm}+mZ_{\pm}}{Z'_{\pm}},
 \quad F=\frac{Z'_+Z'_-}{(Z_+-Z_-)^2},\quad
 Z'^2_{\pm}=d_4+n_1Z_{\pm}+\lambda^2Z_{\pm}^2.\label{19}
 \ea

Let us consider the option:

1a) $\lambda\neq 0.$

The shift $Z_{\pm}:Z_{\pm}\to Z_{\pm}-n_1/2\lambda^2$ in Eq.(\ref{19}) gives:
 \ba
 C_{\pm}=\frac{aZ^2_{\pm}+bZ_{\pm}+c}{Z'_{\pm}},
 \quad F=\frac{Z'_+Z'_-}{(Z_+-Z_-)^2},\nonumber\\
 Z_{\pm}=\sigma_{\pm}\exp(\lambda x_{\pm})+\delta_{\pm}
 \exp(-\lambda x_{\pm}),\quad \sigma_+\delta_+=\sigma_-\delta_-.\label{20}
 \ea
Supposing nonvanishing $\delta_+\neq 0,$ one can write $\sigma_+=\sigma_-\delta_-/\delta_+,$ and inserting $Z_{\pm}$ into
$F,$ one obtains:
 \ba
 F\sim \frac{\delta_-}{(\delta_-\exp(\lambda x_2)-\delta_+\exp(-\lambda
 x_2))^2}-
 \frac{\sigma_-}{(\sigma_-\exp(\lambda x_1)-\delta_+\exp(-\lambda
 x_1))^2}.\label{21}
 \ea
The explicit expressions for superpartner potential, up to a common additive constant, are:
 \ba
 &&V_{1,2}=\frac{a^2}{8\lambda^2}(Z_-^2+Z_+^2)+\frac{a}{4\lambda^2}(b\pm2\lambda^2)
 (Z_-+Z_+)+(a\sigma_-\delta_-+\frac{c}{4})(b\mp 2\lambda^2)\biggl(\frac{Z_-}{Z'^2_-}
 +\frac{Z_+}{Z'^2_+}\biggr)+\nonumber\\&&
 +\frac{1}{8}\biggl((4a\sigma_-\delta_-+c)^2+4\sigma_-\delta_-b(b\mp 4\lambda^2)\biggr)\biggl(\frac{1}{Z'^2_-}+\frac{1}{Z'^2_+}\biggr)+\nonumber\\&&+
 p\biggl(\frac{\delta_-}{(\delta_-\exp(\lambda x_2)-\delta_+\exp(-\lambda
 x_2))^2}+\frac{\sigma_-}{(\sigma_-\exp(\lambda x_1)-\delta_+\exp(-\lambda
 x_1))^2}\biggr), \label{P1}
 \ea
where $Z_{\pm}$ are given in (\ref{20}). We must notice some properties of these potentials. For the particular value
of parameter $a=0$ potentials (\ref{P1}) obey the shape invariance:
\be
V_1(b+4\lambda^2)=V_2(b).\label{shape 1}
\ee
Such potentials were found in \cite{shape} (see Eqs.(44)-(46) there). For some choices of parameters, for example:
$a=c=0, b=4\lambda^2$ or $a=0, b=2\lambda^2, c=4\lambda^2\sqrt{\sigma_-\delta_-}$ one of superpartner potentials $V_1$
is amenable to conventional separation of variables, but the second one is not (see \cite{Ioffe-2}).

The second option is:

1b) $\lambda =0$\quad $n_1\neq 0.$ The shift $Z_{\pm}\to Z_{\pm}-d_4/n_1$ in (\ref{19}) leads to
$Z'^2_{\pm}=n_1Z_{\pm},$ and therefore $Z_{\pm}=n_1x_+^2/4.$ Substitution into expressions for
 $C_{\pm}$ and $F$ gives after redefinition of constants:
 \ba
 Z_{\pm}=n_1x_{\pm}^2/4,\quad
 F_1(2x)=-F_2(2x)=\frac{q}{x^2},\quad
 C_{\pm}=q_3x^3_{\pm}+q_2x_{\pm}+\frac{q_1}{x_{\pm}}.\label{22}
 \ea
Corresponding supersymmetrical partner potentials are:
 \ba
 &&V_{1,2}=\frac{q_3^2}{8}(x_-^6+x_+^6)+
 \frac{q_2q_3}{4}(x_-^4+x_+^4)+\frac{1}{8}(q_2^2+2q_1q_3\pm
 12q_3)(x_-^2+x_+^2)+\nonumber\\&&+\frac{1}{8}(q_1^2\mp
 4q_1)\biggl(\frac{1}{x_-^2}+\frac{1}{x_+^2}\biggr)-\frac{q}{4}
 \biggl(\frac{1}{x_1^2}+\frac{1}{x_2^2}\biggr)+\frac{1}{4}(q_1q_2\pm 4q_2).\label{P2}
 \ea
One can check that the functions $C_{\pm}$ from (\ref{20}) and (\ref{22}) do not satisfy the relation (\ref{redius}), and therefore,
they differ from the earlier of \cite{IGV}.
The case $n_1=\lambda =0$ is not new: it corresponds to Eq.(18) in \cite{Classical}, Eq.(24) in \cite{Ioffe}.

\textbf{2) Polynomial $P_4(U)$ has no degenerate roots.}

One has again two options:

2a) $d_4=0.$

Without loss of generality, the constant $d_3$ can be chosen as $d_3=4.$
After the constant shift $U_{\pm}\to U_{\pm}-d_2/12,$ one obtains the differential equations of the form:
$U'^2_{\pm}=4U^3_{\pm}-g_2U_{\pm}-g_3.$ Their solutions are the well known \cite{bateman} Weierstrass functions $\wp(x_{\pm}).$ Taking into account
the constant shift above, the solution is:
$$U_{\pm}(x_{\pm})=\wp(x_{\pm})-d_2/12.$$
Correspondingly, other coefficient functions are:
 \ba
 C_{\pm}(x_{\pm})=\frac{a+b\wp(x_{\pm})}{\wp'(x_{\pm})}, \quad
 F=\frac{\wp'(x_+)\wp'(x_-)}{(\wp(x_+)-\wp(x_-))^2}=\wp(2x_1)-\wp(2x_2).\label{24}
 \ea

2b) $d_4\neq 0.$ By means of broken-linear transformation
$U_{\pm}=(\alpha Z_{\pm}+\beta)/(\gamma Z_{\pm}+\rho)$
the differential equations
$U'^2_{\pm}=P_4(U_{\pm})$
can be rewritten in the form:
$Z'^2_{\pm}=4Z_{\pm}^3-g_2Z_{\pm}-g_3,$
i.e.
$Z_{\pm}=\wp(x_{\pm})$ are again the Weierstrass functions.
Other coefficient functions are:
 \ba
 C_{\pm}(x_{\pm})=\frac{a\wp^2(x_{\pm})+b\wp(x_{\pm})+c}{\wp'(x_{\pm})},\quad
 F=\frac{U'_+U'_-}{(U_+-U_-)^2}\sim \frac{Z'_+Z'_-}{(Z_+-Z_-)^2}=
 \wp(2x_1)-\wp(2x_2).\label{25}
 \ea
 It is clear that the option 2a) is a particular case of the option 2b).

The superpartner potentials are:
 \ba
 &&V_{1,2}=\frac{a(a\pm 8)}{32}(\wp(x_-)+\wp(x_+))+\frac{1}{32}\sum_{i=1}^3A_i(A_i\mp 8)
 (\wp(x_++\omega_i)+\wp(x_-+\omega_i))+\nonumber\\&&+n(\wp(2x_1)+\wp(2x_2)),\label{P4}
 \ea
where $\omega_1=\omega, \omega_2=\omega +\omega', \omega_3=\omega',$ and $\omega, \omega'$ - semiperiods of
Weierstrass function, and functions
 \ba
 A_i(a,b,c)&=&\frac{c+be_i+ae_i^2}{H_i^2}
 \label{4.1}
 \ea
are expressed in terms of
\be
H_l^2=(e_l-e_m)(e_l-e_n)=2e^2_l+\frac{g_3}{4e_l}=3e^2_l-\frac{g_2}{4},
\quad (l,m,n=1,2,3,\quad l\neq m, l\neq n, m\neq n).\nonumber
\ee
($e_i$ are the roots of cubic equation $4e^3_i-g_2e_i-g_3=0$).
Some properties \cite{bateman} of Weierstrass functions and their roots, such as:
\ba
&&\wp(x+\omega_{\alpha})=e_{\alpha}+\frac{H^2_{\alpha}}{\wp(x)-e_{\alpha}}, \quad \alpha=1,2,3;\nonumber\\
&&\quad \sum_{i=1}^3\frac{e_i}{H^2_i}=0.\nonumber
\ea
were used above in derivation of the explicit expressions (\ref{P4}).

\section*{\normalsize\bf 4. \quad Conclusions.}

The supersymmetric approach has established itself as a powerful method of study of different problems in modern
Quantum Mechanics \cite{review} - \cite{Ioffe-2}. In the case of two-dimensional nonlinear SUSY QM, the general solution of
SUSY intertwining relations seems to be too difficult task. But the same problem with a particular form of supercharges - hyperbolic
metrics in second order part - was rather promising. Until now, even for this metrics only a part of solutions was
built \cite{TMF}, \cite{Classical}, \cite{Ioffe}, \cite{IV} - \cite{IGV}, and the present paper just complete the task providing with
general solution of intertwining relations. It was shown that two solutions (\ref{P2}) and (\ref{P4}) were earlier missed. It is
necessary to mention also that, similarly to previously built systems, these new two-dimensional solutions have the property of
integrability: the symmetry operators $R_{1}=Q^+Q^-$ and $R_2=Q^-Q^+$ being of fourth order in momenta commute with Hamiltonians $H_{1,2}.$
Given the lack of any general method of constructing of integrable models, this result shows once again
the possibilities of SUSY approach in Quantum Mechanics.

\section*{\normalsize\bf \quad Appendix.}

Let us solve the functional-differential equation (\ref{6}):
 \ba
 &&L(A_--A_+)(A_--A_+)[M_+''(A_+)-M_-''(A_-)]+2L''(A_--A_+)(A_--A_+)[M_+(A_+)-M_-(A_-)]-\nonumber\\&&-
 3L'(A_--A_+)[M_+'(A_+)+M_-'(A_-)]=0.\label{A1}
 \ea
 following the procedure of \cite{inozem}. At first, new variables $\tau=(A_--A_+)/2, \rho=-(A_-+A_+)/2$ and
 new functions $M_{\pm}(A_{\pm}),\, \eta(\tau)$:
 \ba
 M_-(A_-)=\frac{\partial}{\partial\tau}N(\tau -\rho),\quad
 M_+(A_+)=\frac{\partial}{\partial\tau}K(\tau +\rho),\quad
 L(A_--A_+)=[\frac{\partial\eta(\tau)}{\partial{\tau}}]^{-2}.\label{A2}
 \ea
 are defined. Then, Eq.(\ref{A1}) reads:
 \ba
 (K-N)'''+(3{\eta'}^{-2}{\eta''}^2-{\eta'}^{-1}\eta''')(K-N)'-
 3{\eta'}^{-1}\eta''(K-N)''=0 \label{A3}
 \ea
(here and below the prime means derivatives over $\tau $ only). One more new function $M,$ defined via $(K-N)'\equiv M\eta',$
allows to rewrite (\ref{A3}) in the form:
$M''\eta'-M\eta''=0,$ i.e.
 \ba
 ({M'}/{\eta'})'=0.\nonumber
 \ea
 Thus the functional equation is obtained:
 \ba
 K(\tau +\rho)-N(\tau
 -\rho)=c_1(\rho)\eta^2(\tau)+c_2(\rho)\eta(\tau)+c_3(\rho),\label{A4}
 \ea
with arbitrary functions $c_k(\rho).$ Both sides of (\ref{A4}) can be
expanded in powers of $\rho ,$ resulting in a system of differential equations:
 \ba
 K(\tau)-N(\tau)=c_1(0)\eta^2(\tau)+c_2(0)\eta(\tau)+c_3(0),\nonumber\\
 (d/d\tau)^k[K(\tau)-(-1)^kN(\tau)]=c_1^{(k)}(0)\eta^2(\eta)+
 c_2^{(k)}\eta(\tau)+c_3^{(k)}(0).\label{A5}
 \ea
It includes three equations:
 \ba
 K(\tau)-N(\tau)=c_1(0)\eta^2(\tau)+c_2(0)\eta(\tau)+c_3(0),\label{A6.1}\\
 (K(\tau)-N(\tau))''=c''_1(0)\eta^2(\tau)+c''_2(0)\eta(\tau)+c''_3(0),\label{A6.2}\\
 (K(\tau)-N(\tau))^{(IV)}=c^{(IV)}_1(0)\eta^2(\tau)+c^{(IV)}_2(0)\eta(\tau)+c^{(IV)}_3(0).\label{A6.3}
 \ea
From Eqs.(\ref{A6.1})-(\ref{A6.2}) and Eqs.(\ref{A6.2})-(\ref{A6.3}) we obtain the system for $\eta'^2(\tau)$ and $\eta''(\tau)$:
 \ba
 c_1(0)(\eta^2(\tau))''+c_2(0)\eta''(\tau)=c''_1(0)\eta^2(\tau)+c''_2(0)
 \eta(\tau)+c''_3(0),\label{A7.1}\\
 c''_1(0)(\eta^2(\tau))''+c''_2(0)\eta''(\tau)=c^{(IV)}_1(0)\eta^2(\tau)+c^{(IV)}_2(0)
 \eta(\tau)+c^{(IV)}_3(0).\label{A7.2}
 \ea
Taking into account $(\eta^2(\tau))''=2(\eta'^2(\tau)+\eta(\tau)\eta''(\tau)),$ we have:
 \ba
 d\eta''(\tau)=e\eta^2(\tau)+a_1\eta(\tau)+b_1,\label{A8.1}\\
 -d\eta'^2(\tau)=e\eta^3(\tau)+a_2\eta^2(\tau)+b_2\eta(\tau)+b_3,\label{A8.2}
 \ea
where $d, e, a_1, b_1, a_2, b_2, b_3$ are arbitrary constants.

At first, we consider the case $d\neq 0.$ After substitution of $\eta''(\tau)$ (obtained from derivative of Eq.(\ref{A8.2})) into Eq.(\ref{A8.1}),
we find that $e=0,$ i.e. to satisfy the system (\ref{A5}) the function $\eta$ must solve the following equation:
 \ba
 {\eta'}^2=a\eta^2+2b\eta +c,\label{A8}
 \ea
where $a,b,c$ - arbitrary new constants, which can be expressed in terms of old constants.

The last case, we have to consider, is $d=c_2(0)c''_1(0)-c''_2(0)c_1(0)=0.$
The value $c''_2(0),$ obtained from the latter relation, has to be inserted into (\ref{A7.1}), leading to
equation:
 \ba
 c_1(0)(c_1(0)\eta^2+c_2(0)\eta)''=c''_1(0)(c_1(0)\eta^2+c_2(0)
 \eta)+c_1(0)c''_3(0).\label{A9}
 \ea
Since $\eta$ was defined up to an additive constant, (\ref{A9}) gives:
 \ba
 {\eta^2}''=4d_2\eta^2+2d_0,\label{A10}
 \ea
or equivalently,
 \ba
 {\eta'}^2=d_2\eta^2+d_0+d_1\eta^{-2}.\label{A11}
 \ea

Thus, in order to satisfy the functional equation (\ref{A4}), one
has to satisfy equations (\ref{A8}) and (\ref{A11}).
The second derivatives of the r.h.s. of (\ref{A4}) over $\tau$ and $\rho$
must be equal, providing the expressions for $c_k(\rho).$
When $\eta$ satisfies Eq.(\ref{A8}), we have the system of differential equations:
  \ba
  c_1''(\rho)=4ac_1(\rho),\quad
  c_2''(\rho)-ac_2(\rho)=6bc_1(\rho),\quad
  c_3''(\rho)=2cc_1(\rho)+bc_2(\rho).\label{A12}
  \ea
For $\eta$ from (\ref{A11}), we obtain:
  \ba
  c_2(\rho)=0,\quad c_1''(\rho)=4d_2c_1(\rho),\quad
  c_3''(\rho)=2d_0c_1(\rho).\label{A13}
  \ea
The solutions of these systems of equations have to be inserted into (\ref{A4}) which provide functions
$K$ and $N.$ Finally, we find functions $L(A_--A_+), M_{\pm}$ and, according to (\ref{5}), equations (\ref{7})-(\ref{13}) for $A_{\pm}.$

\end{document}